# Magneto-ionic control of topological transport in $SrRuO_3$ via band topology engineering

*Xuanchi Zhou [1, 2 *], Xiaohui Yao [1], Xiaomei Qiao [1], Guowei Zhou [1, 2*], Wenjing Huo [1], Shuang Li [1], Huihui Ji [1, 2], Xiaohong Xu [1, 2 *]*

[1] *Key Laboratory of Magnetic Molecules and Magnetic Information Materials of Ministry of Education & School of Materials Science and Engineering, Shanxi Normal University, Taiyuan, 030031, China*

[2] *Research Institute of Materials Science, Shanxi Key Laboratory of Advanced Magnetic Materials and Devices, Shanxi Normal University, Taiyuan 030031, China*

*Authors to whom correspondence should be addressed: *xuanchizhou@sxnu.edu.cn (X. Zhou)*, *zhougw@sxnu.edu.cn (G. Zhou)*, and *xuxh@sxnu.edu.cn (X. Xu)*.

**Abstract**

The interplay between spin-orbit coupling (SOC) and nontrivial band topology in ferromagnets gives rise to a rich landscape of topological transport phenomena such as anomalous Hall effect (AHE) and topological Hall effect (THE). One central goal in modern spintronics lies in the realization of the active control over topological transport phenomena in a reversible fashion, while unambiguously disentangling respective contributions of THE and AHE to the net Hall effect remains a formidable challenge. Here we establish magneto-ionic control as a powerful paradigm for dynamically engineering topological transports in a 4*d*-orbital $SrRuO_3$ system with sizable SOC and itinerant ferromagnetism. Harnessing controllable protonation or oxygen-vacancy incorporation, the Fermi-level upshift relative to avoided band crossings are realized through band-filling control, giving rise to tunable reversal temperature of AHE polarity. Of particular note is the emergence of hump-like Hall anomalies through extensive ionic doping that can be reversibly switched, irrespective of AHE polarity, providing evidence for a THE signal driven by broken inversion symmetry rather than a two-channel AHE. Our findings provide a viable tuning knob for Berry-curvature engineering, enabling on-demand control of topological transports in strong-SOC ferromagnets for low-power, reconfigurable all-oxide spintronic devices.

**Key words**: Oxide spintronics, Anomalous Hall effect, Spin-orbit coupling, Berry curvature, Ionic doping, Topological Hall effect;

## 1. Introduction

The complicate interactions between magnetism and spin-orbit coupling (SOC) underpin exotic topological transport phenomena in magnetic material systems, such as anomalous Hall effect (AHE),[1, 2] topological Hall effect (THE),[3, 4] nonlinear Hall effect [5-7] and spin-orbit torque.[8-11] In a system with broken inversion symmetry, the intrinsic AHE is governed by the integration of Berry curvature over the Brillouin zone.[12, 13] In particular, a strong SOC in ferromagnets introduces nodal points or nodal lines in the spin-polarized bands, giving rise to avoided band crossing near the Fermi level ($E_F$) through SOC-induced gap opening and degeneracy lifting.[14, 15] Such nodal structures with nontrivial topology result in an extensively enhanced Berry curvature that contributes to the intrinsic AHE via emergent magnetic monopoles in momentum space (*k*-space).[16] The polarity and magnitude of AHE are meanwhile determined by the band topology, where different nodes can contribute Berry curvature with opposite signs.[17] Beyond the AHE originating from *k*-space Berry curvature, the Berry phase in real space, arising from a fictitious magnetic field generated by noncoplanar spin textures, can induce a THE in spin-polarized itinerant electrons.[18-20] Nevertheless, the underlying physical origin behind hump-like Hall anomalies in magnetic systems remains a long-standing debate: whether they arise from an intrinsic non-coplanar spin texture [21-26] or from an extrinsic AHE superposition of different magnetic phases with distinct saturation magnetizations and coercivities.[27-30]

Perovskite 4*d*-orbital $SrRuO_3$ exhibits an itinerant ferromagnetism with a Curie temperature ($T_c$) of ~150 K and robust perpendicular magnetic anisotropy (PMA), attractive for engineering scalable all-oxide spintronic devices.[31-33] The critically balanced SOC and electron correlations in $SrRuO_3$ system foster complex topological transports, the AHE of which primarily originates from the intrinsic contribution from *k*-space Berry curvature.[34, 35] The complex competition between different *k*-space Berry curvature sources, dictated by multi-nodal band topology in $SrRuO_3$, results in the sign-reversal AHE, tunable by temperature variation, film thickness, octahedral tilting, and chemical doping.[36] Specifically, temperature reduction has been widely reported to lower the $E_F$ of $SrRuO_3$ with respect to the avoided band-crossings in the $t_{2g}$ orbital of Ru-4*d* band, leading to a sign reversal of AHE polarity from anticlockwise (positive) to clockwise (negative). The correlation between Berry curvature hot spots and AHE polarity provides a fertile ground to flexible design Berry-curvature-driven AHE for reconfigurable spintronic devices. However, the entanglement between *k*-space Berry curvature and real-space Berry phase in $SrRuO_3$ system fundamentally hinders an unambiguous identification of underneath physical picture behind bumps and dips in Hall resistance curves, leaving a debate centered on either a THE or a two-channel AHE.[37-40]

Incorporating mobile ionic species (e.g., proton and oxygen vacancy) into the interstitial sites of transition metal oxides offers a viable pathway for tailoring magnetoelectric transports in a more reversible and flexible fashion through adjusting the ion-electron-lattice interplay.[41-43] Protonation or oxygen vacancy readily donates electron carriers to unoccupied conduction band states, raising the $E_F$ or even driving electronic orbital reconfiguration, while breaking the inversion symmetry through octahedra tilting.[44] Such the ionic evolution provides a fertile ground for controllably engineering the sign and magnitude of AHE and accessing the THE in $SrRuO_3$ through tuning topological band structure and inversion symmetry.[45, 46] Here, the magneto-ionic control over topological transport behaviors in $SrRuO_3$ system is realized, enabling the precise tuning of the reversal temperature ($T_{rev.}$) at which the AHE polarity changes sign through band-filling control and unlocking emergent THE via broken inversion symmetry. The high mobility of intercalated ionic species with small ionic radius enables reversible recovery of hump-like Hall anomalies, providing new insights into their microscopic physical origin. Beyond that, an anomalous elevation in the magnetic coercive field ($H_c$) by over twofold is observed for $SrRuO_3$ through ionic doping, while intrinsic Berry curvature contribution is enhanced, as revealed by the AHE scaling relation. Our findings showcase the robust capability of ionic evolution in controllably adjusting topological transport behaviors in ferromagnets with sizable SOC leveraging Berry-curvature engineering and broken inversion symmetry.

## 2. Results and discussion

The electronic band structure of $SrRuO_3$ with sizable SOC is threaded by symmetry-protected nodal points, endowing it with nontrivial topology and Berry-curvature-driven topological transport phenomena (Figure 1a). Such the nodal points in $SrRuO_3$ give rise to emergent Berry-curvature hot spots that contribute Berry curvature with opposite signs, engendering a non-monotonic, sign-reversing AHE.[36] Engineering the Berry curvature via adjusting the $E_F$ or band topology offers vast opportunities to design the sign and magnitude of AHE in $SrRuO_3$ system for tunable topological spintronics. The magneto-ionic control over band structure serves as a feasible strategy to reversibly elevate the $E_F$ or drive electronic band reconfiguration through band-filling control,[47] and is expected to manipulate the intrinsic AHE arising from Berry curvature. Analogous to pristine $SrRuO_3$, the hump-like Hall anomalies in $SrRuO_3$ induced by hydrogenation can be interpreted as a transport signature of either a THE related to real-space topological spin texture or a two-channel AHE arising from extrinsic magnetic phase superposition,[45, 46] a key priority for future investigation (Figure 1b).

Motivated by this concept, compressive-distorted $SrRuO_3$ film with *a*-axis lattice constant ($a_{0,\text{ film}}$ = 3.93 Å) is deposited on the *c*-faceted $SrTiO_3$ substrate ($a_{0,\text{ sub.}}$ = 3.905 Å), owing to a small lattice mismatch of ~ -0.46 % (Figure 1c). This understanding is further validated by respective high-resolution transmission electron microscopy (HRTEM) technique in Figures 1d and S1, which clearly shows the epitaxial growth of the $SrRuO_3$ film on the $SrTiO_3$ substrate with the expected crystallographic structures confirmed by fast Fourier transform (FFT). Energy dispersion spectrum (EDS) provides evidence for visualizing a chemically distinct and sharp heterointerface in as-deposited $SrRuO_3/SrTiO_3$ (001) bilayer, exemplified by the spatially separated distribution of Ru and Ti elements across the interface (Figure 1e). In addition, the as-grown $SrRuO_3$ film exhibits a relatively smooth surface with terrace edges characteristic of step-flow growth, and its thickness is estimated to be approximately 25 nm (Figure S2). Further consistency in epitaxial relationship of $SrRuO_3/SrTiO_3$ heterostructure is collaborated by reciprocal space mapping (RSM) around the $SrTiO_3$ (103) and (013) reflections (Figure 1f), where identical *in-plane* vector (e.g., $Q_{\parallel}$) demonstrates the epitaxial growth. An enlarged *out-of-plane* vector (e.g., $Q_{\perp}$) for $SrTiO_3$ substrate relative to $SrRuO_3$ film verifies an *in-plane* compression distortion of $SrRuO_3$. In particular, the orthorhombic structure of the grown $SrRuO_3$ film is identified by distinct $Q_{\parallel}$ values for the {103} diffraction peaks, attributed to the $RuO_6$ octahedral tilting, differing from invariant peak positions characteristic of a tetragonal film.[45] Furthermore, the ferromagnetic ordering is observed for the deposited $SrRuO_3$ film below the $T_c$ of ~ 130 K, and the temperature at which the kink appears in its itinerant electrical transport behavior ($T_{kink}$) coincides with $T_c$ (Figure S3).[48]

To realize the effective incorporation of hydrogen into $SrRuO_3$ lattice, Pt dots as catalyst are sputtered onto the surface of $SrRuO_3$ film that facilitates the dissociation from $H_2$ molecules into protons and electrons, a process known as the hydrogen spillover strategy.[49] Assisted by thermal annealing ranging from 70 ºC to 200 ºC with a period of 1 h, hydrogen-triggered lattice expansion of $SrRuO_3$ along the *out-of-plane* direction is observed without perturbing expected lattice framework, as evidenced by their X-ray diffraction (XRD) patterns (Figures 1g and S4). It is worth noting that raising the hydrogenation temperature (70 ºC, 150 ºC and 200 ºC) is prone to result in a harsher hydrogenation degree (denoted as $H_{x1}SrRuO_3$, $H_{x2}SrRuO_3$ and $H_{x3}SrRuO_3$), whereas excessively high temperatures can disrupt the itinerant ferromagnetism of $SrRuO_3$ or even trigger redox reactions.[45] Consistent with prior reports,[46] protonation reduces the magnitudes of both the $T_c$ and $T_{kink}$ in $SrRuO_3$ film, while simultaneously elevating its material resistivity (Figure S5).

Temperature-driven sign reversal of AHE polarity in 25 nm-thick $SrRuO_3$ is demonstrated by the Hall resistance curves ($R_{xy}$-$H$) in Figure 2a, where cooling to the $T_{rev.}$ of 110 K results in a sign change from anticlockwise (positive) to clockwise (negative). Such the temperature-controlled sign reversal of AHE in $SrRuO_3$ system is associated with a $E_F$ downshift relative to the band-crossing points as the temperature decreases.[36] The hydrogenation of $SrRuO_3$ gradually reduces the $T_{rev.}$ from 110 K to 83 K, as their $\rho_{xy}$-$T$ tendencies shown in Figure 2b. In addition, protonation triggers an increase in the $H_c$ of $SrRuO_3$, with a rapid rise upon cooling due to domain-wall pinning induced by hydrogen doping (Figure S6). Notably, the hydrogen-induced decrease in the $T_{rev.}$ of $SrRuO_3$ is readily reversible toward the pristine state through dehydrogenation using oxidative annealing or exposure to the air, albeit with complete restoration impeded by residual hydrogen trapped in deeper layers (Figures 2c and S7). The reversible magneto-ionic control over topological transport behavior of $SrRuO_3$, enabled by the high mobility of incorporated hydrogens,[50] benefits the engineering of all-oxide spintronic device. However, the hydrogenated state of $SrRuO_3$ remains relatively stable for the first 11 days of ambient exposure, but undergoes a gradual recovery after 15 days, as demonstrated by the similar AHE in Figure S8.

It is particularly worth noting that hump-like Hall anomalies emerge in $H_{x3}SrRuO_3$ system at 50 K under extensive hydrogenation, where a characteristic hump or dip is detected near the coercive field, superimposed on the AHE loop. In addition, hump-like Hall anomalies are observed only for $H_{x3}SrRuO_3$ under extensive hydrogenation, in stark contrast to their $H_{x1}SrRuO_3$ and $H_{x2}SrRuO_3$ counterpart. The hump-like Hall anomalies are attributed to either 1) a THE induced by the broken inversion symmetry, or 2) a two-channel AHE arising from the coexistence of pristine and hydrogenated phases. Remarkably, such hump-like Hall anomalies in hydrogenated $SrRuO_3$ are ultimately vanished upon sequential oxidative annealing (Figure 2d). Given that a lower $T_{rev.}$ (~100 K) of $SrRuO_3$ after dehydrogenation compared to the pristine state (~110 K), hump-like Hall anomalies reversibly disappear in dehydrogenated $SrRuO_3$, despite the coexistence of hydrogenated and pristine phases. Therefore, the hump-like Hall anomalies are more likely a consequence of the intrinsic THE induced by broken inversion symmetry, rather than the superposition of pristine and hydrogenated magnetic phases. Hydrogen-triggered broken inversion symmetry may be associated with the $RuO_6$ octahedral tilting or a polarization field triggered by hydrogen concentration gradient, worthy of further investigation. The octahedral tilting of $SrRuO_3$ can indirectly reflected by the elevated material resistivity and reduced $T_c$ through hydrogenation.[45, 51]

To identify the effective incorporation of hydrogens, time-of-flight secondary ion mass spectrometry (ToF-SIMS) analysis is performed, where a $Cs^+$ ion beam sputters the $SrRuO_3$ lattice and the sputtering-time-dependent intensity of ejected molecular ions reveals the depth profile of incorporated hydrogens (Figure 3a). The presence of hydrogens is uncovered by three-dimensional ToF-SIMS element maps for $SrRuO_3/SrTiO_3$ (001) heterostructure, in which a more pronounced H signal is detected for the $SrRuO_3$ film region with respect to the $SrTiO_3$ substrate. This finding is in agreement with the depth profile of elementary distribution in Figure 3b, where a higher hydrogen intensity is observed for the $SrRuO_3$ film region. The two-dimensional ToF-SIMS element maps unveils a relatively homogenous hydrogen distribution within the lattice of $SrRuO_3$ (Figure S9). In addition, the hydrogen intensity as detected by ToF-SIMS analyzer gradually reduces away from the surface of $SrRuO_3$ film, following the Fick's diffusion law. Such the hydrogen concentration gradient in $SrRuO_3$, vertical to the heterointerface, may induce a built-in polarization field that breaks the inversion symmetry to drive the emergence of THE.[46]

The hydrogen incorporation triggers the anisotropy in the lattice expansion of $SrRuO_3$ along the *out-of-plane* direction, as demonstrated by a reduced $Q_\perp$ from 7.61 $Å^{-1}$ to 7.56 $Å^{-1}$, while the *in-plane* lattice remains clamped by the epitaxial template (Figure 3c). Benefiting from the AHE scaling relation as described by $\sigma_{xy} = A\sigma_{xx}^{1.6}$+B,[52] hydrogenation is prone to enlarge the intrinsic contribution from Berry curvature (coefficient B) (Figures 3d and S10), in comparison with scattering-dependent extrinsic contribution arising from the skew scattering or side jump ($A\sigma_{xx}^{1.6}$).

Hydrogen doping depresses the hybridization of Ru-3*d* and O-2*p* orbitals, as indicated by the decreased peak A intensity in the O-*K* edge of soft X-ray absorption spectroscopy (sXAS) spectra, similar to hydrogenated $SrCoO_{2.5}$, $NiCo_2O_4$ and $La_{1-x}Sr_xMnO_3$ (Figure 3e).[53, 54] Therefore, hydrogen-related band-filling control elevates the $E_F$ relative to avoided band-crossing points, reducing the $T_{rev.}$ of $SrRuO_3$ in a reversible fashion, meanwhile driving the emergence of THE via the broken inversion symmetry.

In order to generalize the ionic evolution strategy, similar electron doping introduced by oxygen vacancies are exploited to manipulate topological transport behaviors of $SrRuO_3$. To introduce oxygen deficiency into the $SrRuO_3$ lattice, two representative strategies are utilized: 1) high-vacuum annealing at temperatures ranging from 300 ºC to 400 ºC for 1 h;[55] 2) directional oxygen ionic transport via oxygen chemical potential mismatch ($\Delta\mu_O$) (Figure 4a).[56, 57] High-temperature annealing under a vacuum atmosphere ($<10^{-5}$ Pa) facilitates the oxygen desorption from the $SrRuO_3$ lattice, with the degree increasing at higher temperatures, denoted as $SrRuO_{3-x1}$, $SrRuO_{3-x2}$ and $SrRuO_{3-x3}$. The enlarged $\mu_O$ for $SrRuO_3$ in comparison with $SrTiO_3$ allows for direction oxygen ionic transport across the heterointerface from $SrRuO_3$ to $SrTiO_3$, thereby inducing oxygen deficiency in $SrRuO_3$ underlayer. In addition, a 2.5 nm-thick oxygen-deficient $SrTiO_3$ overlayer, deposited at 250 ºC as an oxygen reservoir, is critical for accommodating the transferred oxygen ions (Figure S11), the amorphization of which is visualized by HR-TEM images in Figure 4b. The chemically sharp interface between $SrTiO_3$ overlayer, $SrRuO_3$ interlayer and $SrTiO_3$ substrate is demonstrated by respective EDS mappings (Figures 4c and S12). Moreover, the introduction of $SrTiO_3$ capping layer cannot deteriorate the surface roughness of the grown $SrRuO_3$ film (Figure S13).

Oxygen deficiency introduced by either vacuum annealing or $\Delta\mu_O$-driven oxygen ionic migration simultaneously decreases both the $T_c$ and $T_{kink}$ of $SrRuO_3$, accompanied by an increase in the material resistivity (Figures 4d and S14). Beyond magnetoelectric phase modulations, oxygen vacancies trigger the lattice expansion of $SrRuO_3$, as evidenced by the leftward shift of $SrRuO_3$ (002) (Figure S15). Such the defect-mediated lattice expansion of $SrRuO_3$ becomes more pronounced as the vacuum-annealing temperature increases, and in all cases exceeds that induced by $\Delta\mu_O$-driven oxygen ionic migration. This observation is consistent with the modulation on the $T_{rev.}$ of AHE through oxygen defects, wherein vacuum annealing leads to a more pronounced reduction in the $T_{rev.}$, compared to $SrTiO_3/SrRuO_3$ heterostructure (Figures 4e-4f and S16). Defect-driven reduction in the $T_{rev.}$ of $SrRuO_3$ aligns well with the hydrogenated counterpart, unraveling a generalized band-filling control over topological transport behaviors via ionic evolution. This is also related to defect-related band-filling control, where each oxygen vacancy donates two electron carriers into conduction band and suppresses the orbital hybridization (Figure S17). As a result, the $E_F$ upshift of oxygen-deficient $SrRuO_3$ in comparison with the band-crossing points reduces the $T_{rev.}$ of AHE polarity. Notably, the introduction of oxygen vacancies renders an anomalous threefold increase in the $H_c$ of $SrRuO_3$, which rises rapidly as the temperature decreases (Figure 4g). This result is associated with the domain-wall pinning by oxygen defects, in agreement with prior report.[58] Nevertheless, further extending the annealing period to 3 h or 6 h leads to more pronounced structural transformation and magnetoelectric phase modulations (Figures 4h-4i and S18). A substantial introduction of oxygen vacancies to $SrRuO_3$ lattice gives rise to the absence of the AHE sign reversal, which showcases a positive AHE over entire measured temperature range. Analogous to hydrogenation, a hump-like THE signal emerges upon introducing extensive oxygen vacancies (Figure 4j).

Incorporating mobile ionic species (proton and oxygen vacancy) offers a feasible pathway for modulating topological transport behaviors of $SrRuO_3$ with a sizable SOC in a reversible fashion. The intimate relationship between avoided band-crossing points and AHE polarity enables the possibility in facilely adjusting the $T_{rev.}$ of AHE sign in $SrRuO_3$ system through band-filling control (Figure 4k). Ionic-evolution-induced electron doping shifts the $E_F$ upward with respect to the band-crossing points, leading to a reversible decrease in $T_{rev.}$ of the AHE sign. Further reducing the film thickness of $SrRuO_3$ to 4.5 nm offers an alternative pathway for driving the decrease in the $T_{rev.}$ of $SrRuO_3$ from 110 K to 90 K, in comparison with a 25-thick counterpart through modulation of the band topology (Figures 4l and S19). Likewise, unidirectional oxygen ionic migration via the introduction of $SrTiO_3$ capping layer can further reduce $T_{rev.}$ in the 4.5 nm-thick $SrRuO_3$ film from 90 K to 85 K. The ability to regulate the $T_{rev.}$ through ionic evolution is superior to traditional chemical doping, for example, substituting B-site $Ru^{4+}$ with $Mo^{6+}$, which reduces $T_{rev.}$ by only 8 K through an irreversible process.[59] In addition, the domain wall pinning induced by ionic incorporation triggers an anomalous elevation in the $H_c$ of $SrRuO_3$ by 2-3 times.

Hump-like Hall anomalies emerge in the Hall resistance curve of $SrRuO_3$ through extensive protonation or oxygen deficiency. The ongoing controversy over such hump-like Hall anomalies centers on whether their physical origin is a THE or an AHE. Despite an incomplete $T_{rev.}$ recovery via dehydrogenation due to phase coexistence, the reversible disappearance of hump-like Hall anomalies strongly favors a THE interpretation over a two-channel AHE (Figure 3d), providing fundamentally new insights into its physical picture. This understanding is further supported by the hump-like Hall signal that emerge in $SrRuO_3$ upon ionic evolution over a relatively wide temperature range, irrespective of the AHE polarity (Figure 4j). This also contrasts with the previous two-channel AHE picture that magnetic inhomogeneities tend to appear near the boundary of the $T_{rev.}$ of AHE polarity. The above findings reveal a THE-mediated transport behavior in $SrRuO_3$ induced by ionic doping, which calls for further *in-situ* characterization of real-space noncoplanar spin textures, a task that is challenged by their volatile nature. The non-trivial topology associated with avoided band crossing provides a fertile ground for design topological transport phenomena through Berry curvature engineering and inversion-symmetry breaking via ionic evolution.

## 3. Conclusion

In summary, we identify ionic evolution as a powerful tuning knob for adjusting topological transport behaviors in $SrRuO_3$ system with sizable SOC, giving rise to tunable AHE and THE phenomena. Electron-doping-mediated Berry curvature engineering lowers the $T_{rev.}$ of AHE polarity in $SrRuO_3$ by driving a Fermi-level upshift relative to the band-crossing points via ionic doping. The broken inversion symmetry induced by ionic doping triggers emergent hump-like Hall anomalies in $SrRuO_3$ system. Magneto-ionic control over the inversion symmetry of $SrRuO_3$ can be reversibly recovered through dehydrogenation, deepening the understanding of hump-like Hall anomalies under controllable protonation or oxygen-vacancy incorporation, which are ascribed to the THE rather than a two-channel AHE. This work establishes ionic evolution as a versatile platform for Berry-curvature engineering and topological transport control, opening a new avenue toward reconfigurable low-power spintronic devices using ferromagnets with sizable SOC.

## 4. Experimental Section

*Fabrication of the grown $SrRuO_3$ films:* Epitaxial $SrRuO_3$ films were grown on single-crystalline (001)-oriented $SrTiO_3$ (001) substrates using laser molecular beam epitaxy (LMBE). The $SrRuO_3$ film deposition was performed at 700 °C. The oxygen partial pressure, target-to-substrate distance, and laser fluence were optimized as 26 Pa, 42 mm, and 1.0 J $cm^{-2}$, respectively. After the film deposition, the grown $SrRuO_3$ films were subjected to an *in-situ* post-annealing treatment in an oxygen-rich atmosphere at an oxygen partial pressure of $2.6\times10^4$ Pa for 45 minutes to minimize oxygen vacancies and improve the crystallinity. The $SrRuO_3$/$SrTiO_3$ (001) heterostructures were subsequently cooled to room temperature under the identical oxygen partial pressure. Before the hydrogenation, the 20 nm-thick platinum dots were sputtered into the surface of the grown $SrRuO_3$ films via exploiting the magnetron sputtering technique. Finally, based on the hydrogen spillover strategy, as-made Pt/$SrRuO_3$/$SrTiO_3$ (001) heterostructures were annealed in a 5 % $H_2$/Ar forming gas for effectively realizing the hydrogenation. The oxygen-deficient $SrRuO_3$ films were obtained by annealing from 300 ºC to 400 ºC under a high vacuum atmosphere ($P_{O_2} < 10^{-5}$ Pa) or capping with an $SrTiO_3$ overlayer. The $SrTiO_3$ capping layer was grown at 250 °C under an oxygen partial pressure of 0.5 Pa. Thereafter, the as-grown trilayer was cooled to ambient temperature under the same oxygen pressure without intentional adjustment, thereby yielding a $SrTiO_3$/$SrRuO_3$/$SrTiO_3$ (001) trilayer.

*Material characterizations:* The crystal structures of the grown $SrRuO_3$ films were probed via using the X-ray diffraction (XRD) (Rigaku, Ultima IV) and high-resolution transmission electron microscopy (HRTEM) (FEI, Tecnai G2 20 S-TWIN). The epitaxial growth of the grown $SrRuO_3$ film is identified by using the reciprocal space mapping (RSM) (Rigaku, Ultima IV). The surface topology and film thickness of the grown $SrRuO_3$ films is identified by using the atomic force microscope (AFM) (Bruker, Dimension Icon). The electronic structure of $SrRuO_3$ films was further explored through soft X-ray absorption spectroscopy (sXAS) analysis, as conducted at the Shanghai Synchrotron Radiation Facility (SSRF) on beamline BL08U1A and National Synchrotron Radiation Laboratory (NSRL) on beamline BL12B-b. Magnetic properties were measured using a superconducting quantum interference device (SQUID) (Quantum Design). Temperature dependent resistivity and anomalous Hall effect of the deposited $SrRuO_3$ films was measured by using a commercial physical property measurement system (PPMS) (Quantum design). The elementary depth profile is examined by using the time-of-flight secondary ion mass spectrometry (ToF-SIMS) technique (ION-TOF GmbH, TOF.SIMS 5).

## Figures and captions

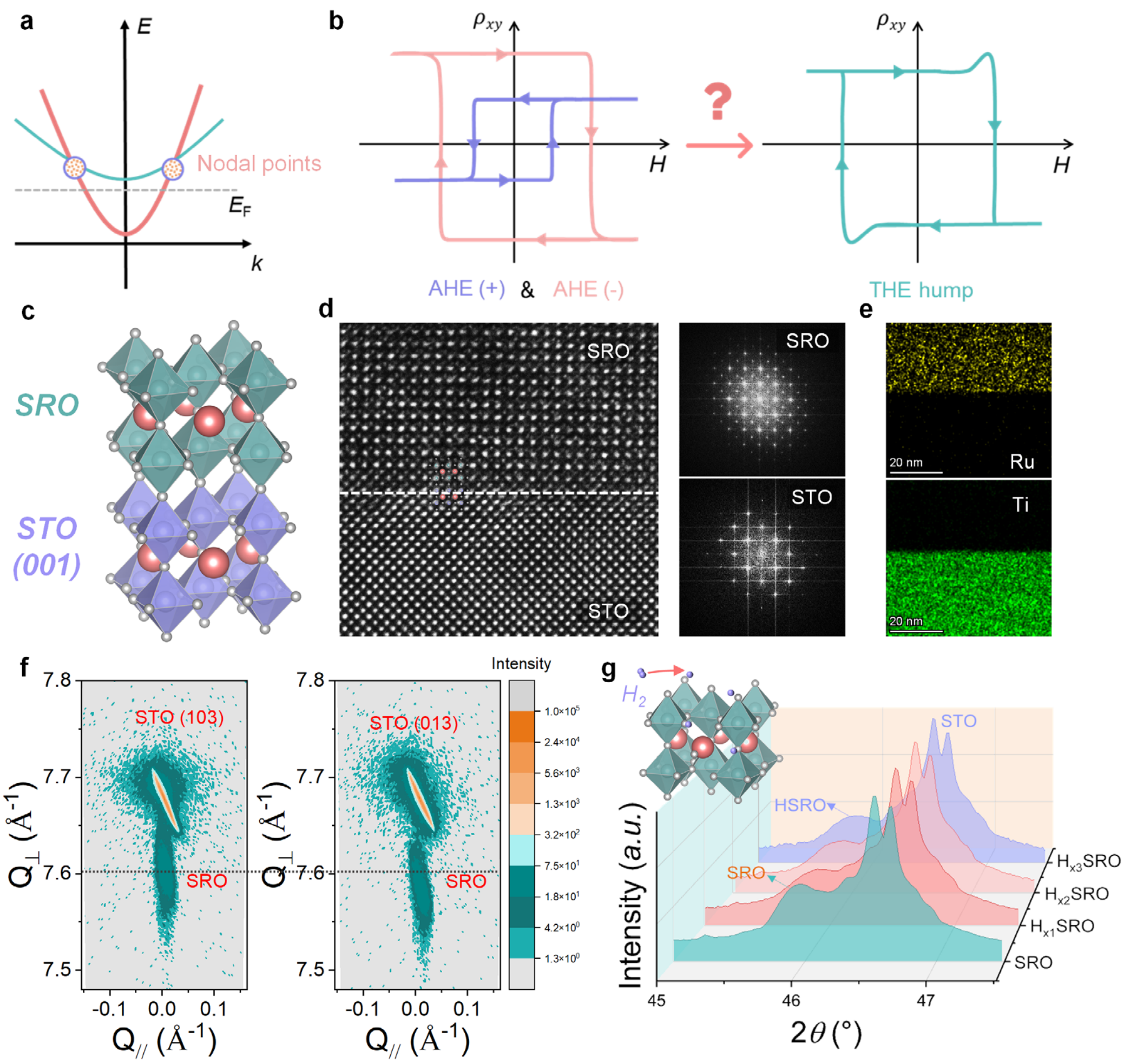

**Figure 1. Hydrogen-driven structural evolution in $SrRuO_3$ system. a,** Schematic of the topological band structure of $SrRuO_3$ (SRO) with nodal points. **b,** Schematic illustration of the controversy over the physical picture of hump-like Hall anomalies in SRO. **c,** Schematic illustration of the lattice framework of the SRO/$SrTiO_3$ (STO) (001) bilayer. **d,** Zoom-in images for cross-sectional high-resolution transmission electron microscopy (HRTEM) and respective Fast Fourier Transform (FFT) images for SRO/STO bilayer. **e.** Ruthenium and titanium elementary distributions for as-grown SRO/STO heterostructure characterized by using energy dispersion spectrum (EDS). **f,** Reciprocal space mapping (RSM) of SRO films around the STO (103) and (013) substrate reflections. **g.** X-ray diffraction (XRD) patterns compared for SRO films through hydrogenation.

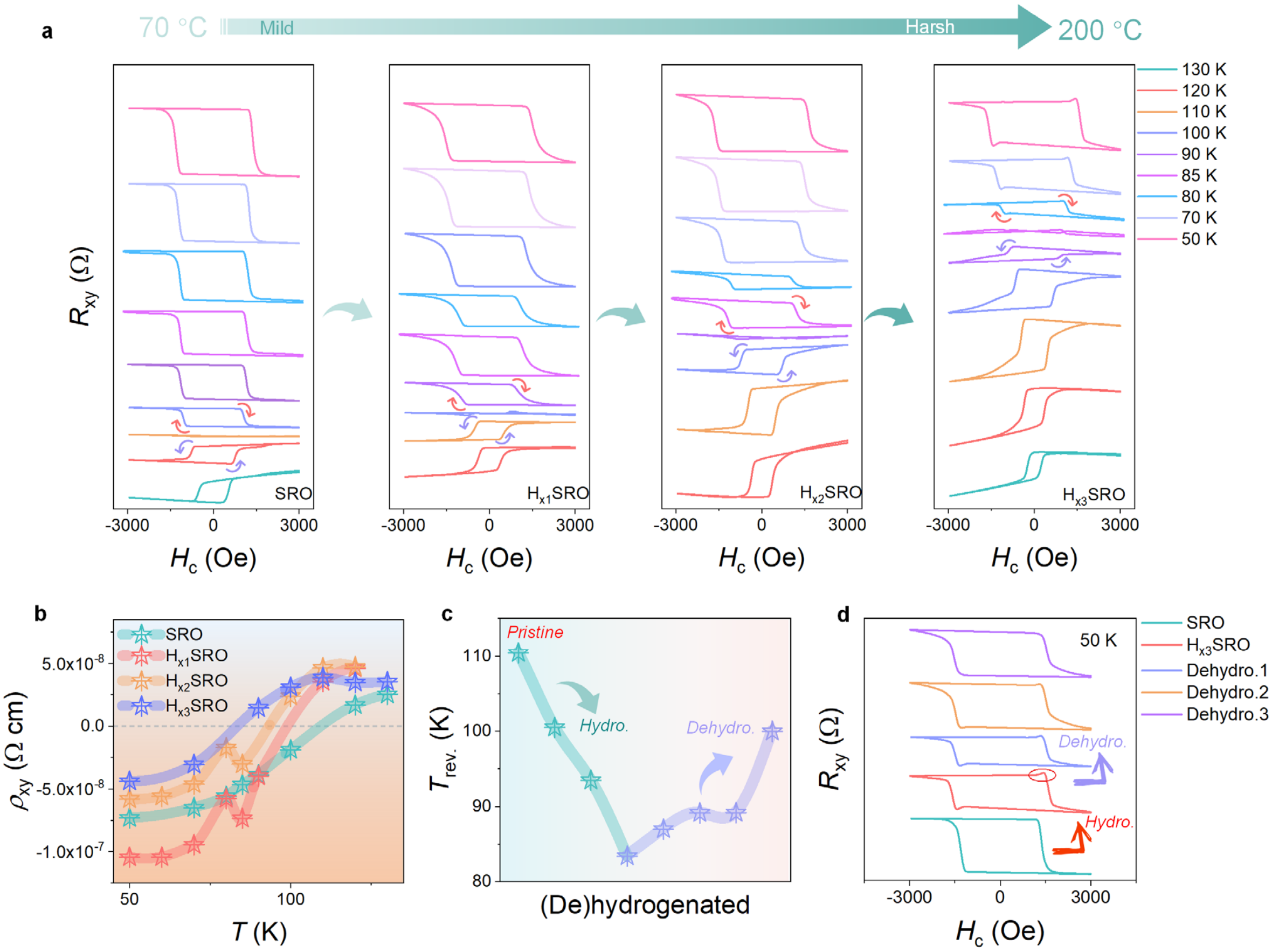

**Figure 2. Manipulating topological transport behaviors through hydrogenation. a,** Hydrogenation-mediated regulation on the anomalous Hall effect (AHE) in SRO films. **b,** Temperature dependence of the anomalous Hall resistivity during the (de)hydrogenated process. **c,** The reversal temperature ($T_{rev.}$) of SRO through (de)hydrogenation. **d,** Emergent hump-like Hall anomalies in the Hall resistance curves of SRO films upon (de)hydrogenation.

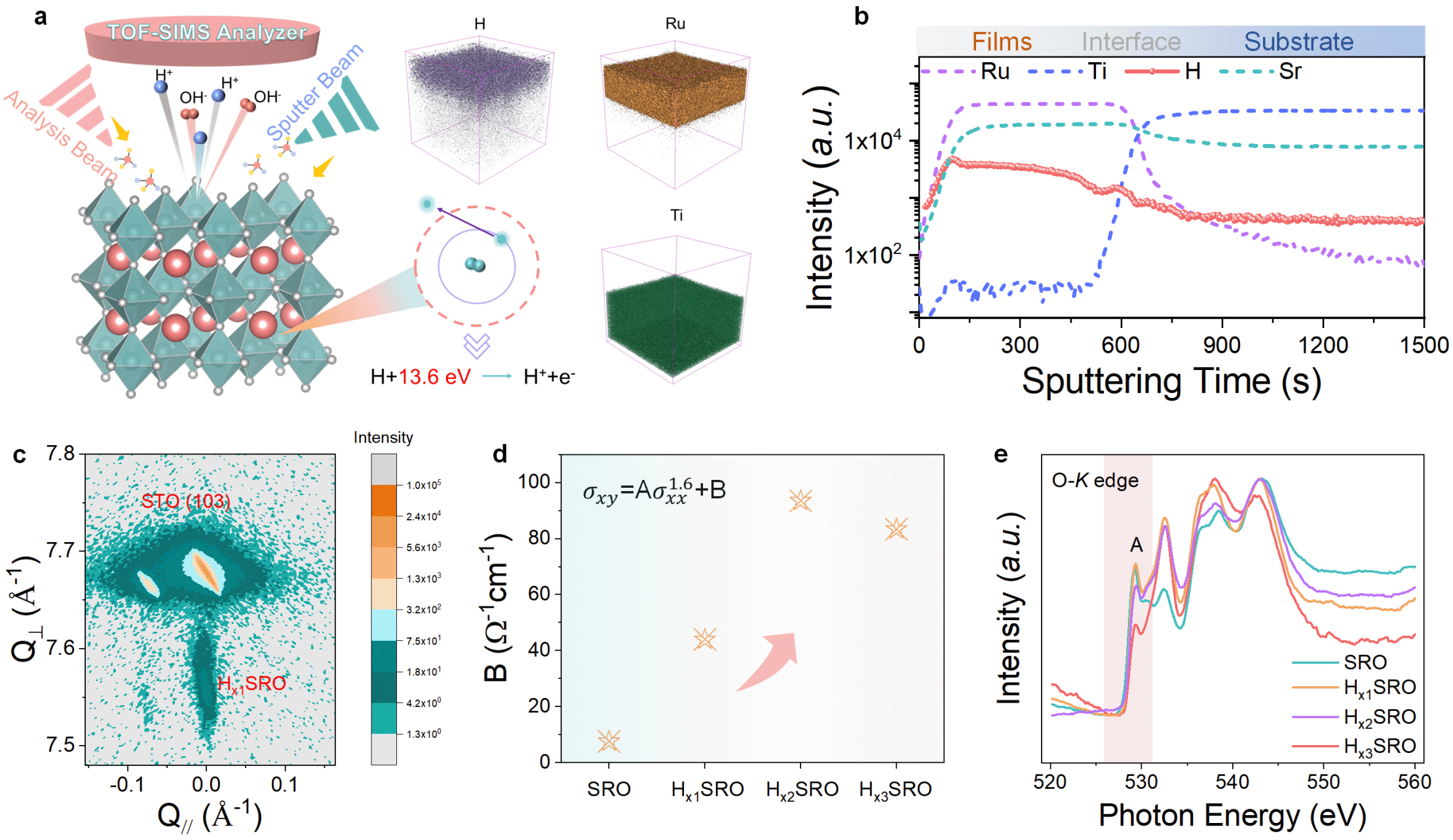

**Figure 3. Berry-curvature-driven topological transports in SRO. a,** Schematic of the principle in time-of-flight secondary ion mass spectrometry (ToF-SIMS) and three-dimensional ToF-SIMS element maps for SRO/STO (001) bilayer. **b,** Depth-profile of the elementary distribution in SRO/STO (001) bilayer. **c,** RSM spectra for hydrogenated SRO/STO heterostructure ($H_{x1}$SRO). **d,** Evolution of the Berry curvature contribution (B value) in hydrogenated SRO thin films via fitting the AHE scaling relation. **e,** Soft X-ray absorption spectroscopy (sXAS) for the O *K*-edge compared for SRO films through hydrogenation.

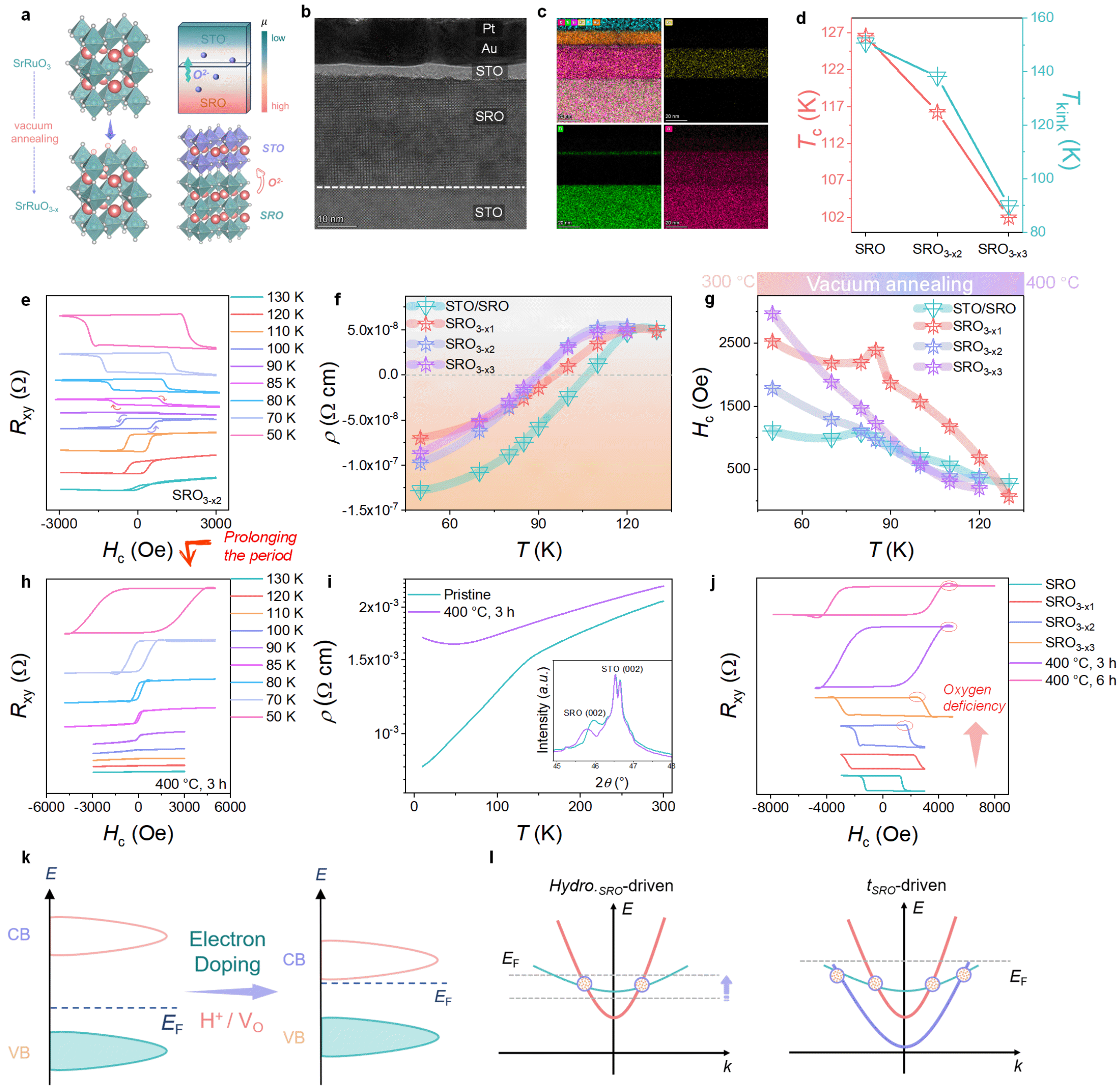

**Figure 4. Defect-engineered magneto-ionic control of topological transport phenomena. a,** Schematic illustration of oxygen ionic transport as driven by vacuum annealing and oxygen chemical potential. **b,** HETEM images of STO/SRO/STO (001) trilayer. **c,** Ruthenium, titanium and oxygen elementary distributions for as-grown STO/SRO/STO (001) trilayer characterized by using EDS. **d,** Evolution of the Curie temperature ($T_c$) and kink temperature ($T_{kink}$) of SRO through oxygen deficiency. **e,** Temperature-dependent AHE in oxygen-deficient SRO ($SRO_{3-x2}$) films. **f-g,** Temperature-dependence of **f,** anomalous Hall resistivity and **g,** magnetic coercive Field ($H_c$) for oxygen-deficient SRO films. **h,** Temperature-dependent AHE for SRO after an extended vacuum annealing period (400 °C, 3 h). **i,** Temperature dependence of material resistivity ($\rho$-$T$) as measured for pristine and oxygen-deficient SRO films. **j,** Hump-like Hall anomalies in Hall resistance curves of oxygen-deficient SRO. **k,** Schematic of band structure modifications induced by electron doping. **l,** Schematic illustration of hydrogen-driven and thickness-driven modifications of the electronic band structure near the Fermi level in SRO.